\documentclass[sigconf]{acmart}

\AtBeginDocument{%
  \providecommand\BibTeX{{%
    \normalfont B\kern-0.5em{\scshape i\kern-0.25em b}\kern-0.8em\TeX}}}

\copyrightyear{2023} 
\acmYear{2023} 
\setcopyright{acmlicensed}\acmConference[IDC '23]{Interaction Design and Children}{June 19--23, 2023}{Chicago, IL, USA}
\acmBooktitle{Interaction Design and Children (IDC '23), June 19--23, 2023, Chicago, IL, USA}
\acmPrice{15.00}
\acmDOI{10.1145/3585088.3589386}
\acmISBN{979-8-4007-0131-3/23/06}

\begin{document}


\title[Family Theories in Child-Robot Interactions]{Family Theories in Child-Robot Interactions: Understanding Families as a Whole for Child-Robot Interaction Design}





\author{Bengisu Cagiltay}
\email{bengisu@cs.wisc.edu}
\affiliation{%
  \institution{Computer Sciences Department, University of Wisconsin-Madison}
  \city{Madison}
  \state{WI}
  \country{USA}
}

\author{Bilge Mutlu}
\email{bilge@cs.wisc.edu}
\affiliation{%
  \institution{Computer Sciences Department, University of Wisconsin-Madison}
  \city{Madison}
  \state{WI}
  \country{USA}
}

\author{Margaret Kerr}
\email{margaret.kerr@wisc.edu}
\affiliation{%
  \institution{Human Development and Family Studies, University of Wisconsin-Madison}
  \city{Madison}
  \state{WI}
  \country{USA}
}

\renewcommand{\shortauthors}{Bengisu Cagiltay, Bilge Mutlu, Margaret Kerr}

\begin{abstract}
In this work, we discuss a theoretically motivated family-centered design approach for child-robot interactions, adapted by Family Systems Theory (FST) and Family Ecological Model (FEM). Long-term engagement and acceptance of robots in the home is influenced by factors that surround the child and the family, such as child-sibling-parent relationships and family routines, rituals, and values. A family-centered approach to interaction design is essential when developing in-home technology for children, especially for social agents like robots with which they can form connections and relationships. We review related literature in family theories and connect it with child-robot interaction and child-computer interaction research. We present two case studies that exemplify how family theories, FST and FEM, can inform the integration of robots into homes, particularly research into child-robot and family-robot interaction. Finally, we pose five overarching recommendations for a family-centered design approach in child-robot interactions. 
\end{abstract}



\begin{teaserfigure}
\centering
  \includegraphics[width=\textwidth]{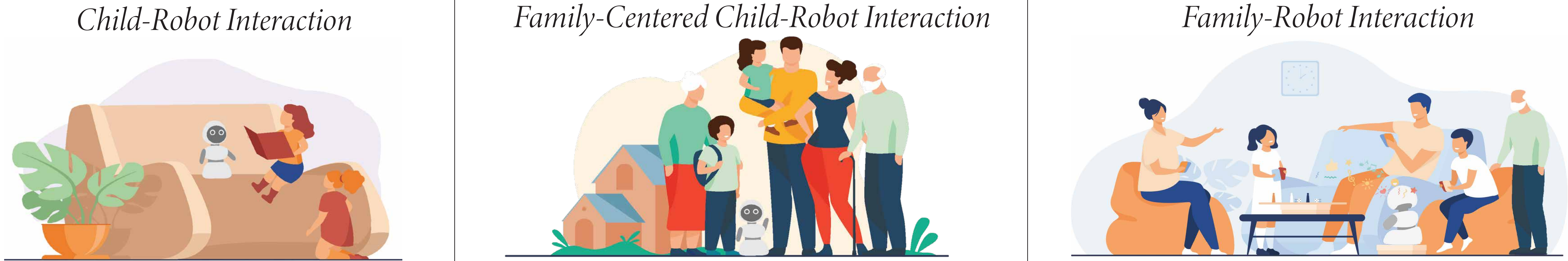}
  \Description{On the left panel, titled "Child-Robot Interaction" there is a child reading to a robot on the couch and another child listening in to the interaction. The child-robot dyad is circled with red dashed lines. On the middle panel, titled "Family Centered Child-Robot Interaction" there is a family with parents, grandparents, two children, and a robot standing, the child-robot dyad is circled with red dashed lines as well as the whole family is circled with the dashed lines. On the right panel, titled "Family-Robot Interaction" there is a family of two children, parents, and a grandfather, sitting in the living room, playing a card game with a robot.}
  \caption{\textit{Conceptual illustrations of child-robot and family-robot interactions:} Child-robot interactions typically capture dyadic relationships between a child and a robot. Family-centered child-robot interactions, however, center the child-robot dyad but also consider the surrounding family members that influence the interaction between the child and the robot. In this paper, we take a theory-motivated approach to discuss family-centered child-robot interaction. We also propose the concept of \textit{family-robot interactions}, where the family as a whole interacts with the robot together in a multi-party interaction setting.}
  \label{fig:teaser}
\end{teaserfigure}

\maketitle

\section{Introduction}

Conversational agents, such as Amazon's Alexa and Google's Home Assistant, and autonomous cleaning robots, such as the Roomba, have become ubiquitous in homes, and daily lives of families, in the past decade. Research into the use of these technologies have found that parents utilize smart speakers to complement and extend their parenting goals, to improve communication among family members, regulate the use of technology, and augment parenting tasks such as setting routines and schedules~\cite{beneteau2020parenting}. As global the market for household robots continues to grow, reaching an estimated USD 21.94 billion by 2027~\cite{reportlinker_2022}, widespread adoption of robots in domestic environments are expected and there is a greater need to study the long-term adoption of social robots in a family-inclusive approach. 

Research in robotics has shown that social robots are effective in delivering interventions for children with autism~\cite{cabibihan2013robots} and are perceived as being less judgemental by children especially in educational contexts when learning math or reading~\cite{smakman2021moral}. 
Social robots can help with anxiety and emotion regulation~\cite{jeong2015social}; support children in learning numerical skills~\cite{ho2021robomath}; facilitate creativity~\cite{alves2016boosting}; and promote prosocial behavior development~\cite{peter2021can}. 
Design-based research shows that children prefer to read with, play with, and take care of a social companion robot over extended periods of time at their homes~\cite{cagiltay2020investigating, cagiltay2022exploring, lee2022unboxing}. 
However, homes include many members, complex dynamics and family structures that impact perspectives affecting the family's long-term engagement and acceptance of the robot. 
In order to fully integrate social robots into the home environment, the design of this technology must consider not just child-robot interactions but \textit{families as a whole}.


In this paper, we propose taking a family-centered approach to designing social robot interactions for children and their families (i.e., family-robot interactions). We review related work in child-robot interaction and child-computer interaction research and reflect on how family theories can inform the design of social robots that will be deployed in the homes of children and their families. 
Furthermore, we discuss the strengths and limitations of a family-centered approach for child-robot interactions. Building on this discussion, we propose five key recommendations for how child-robot interaction researchers can follow a family-centered approach. 
Overall, we discuss how human-robot interaction research can benefit from understanding families from a theoretical point of view to better design in-home social robot interactions for children and families. Specifically, the following question guides this work: \textit{How can family theories, i.e., Family Systems Theory and Family Ecological Model, inform how robots might be integrated into domestic environments to support child-robot and family-robot interactions?}
\section{Related Work}

\subsection{Family Theories}
Early applications of family theories were in family clinical practices, proposed by Bowen's Family Systems Theory (FST)~\cite{bowen1966use}. FST argues that families are a complex social system where members interact to influence each other's behavior. \citet{cox1997families} describes family systems using a biological systems metaphor, e.g., just as cells form organisms, families are formed of smaller subsystems (such as interactions between parents, spouses, or siblings) and are a part of larger systems. In this perspective, a family is greater than its sum of parts, has a hierarchy of subsystems, and is capable of adaptive self-stabilization and self-organization depending on the changes in the larger system. \citet{cox1997families} summarizes this as the following: \textit{``Any individual family member is inextricably embedded in the larger family system and can never be fully understood independent of the context of that system. ''}
From this perspective, a social robot in the home is arguably a member of the household and family system, can form different connections and relationships with other family members, influence the behaviors of family members, and other members of the system can influence the behavior of the robot. A family-systems approach applied in child-robot interaction research can help us understand how a robot can facilitate parent-child, sibling-sibling~\cite{cox2010family}, or intergenerational connections through shared social interactions. Family-systems could also help us understand parental expectations, concerns and acceptance of social robots in the home~\cite{lin2020parental}.

However, families are also surrounded with ecological factors that influence their relationships and interactions. On this note, Brofenbrenner's bioecological theory~\cite{bronfenbrenner1979ecology, bronfenbrenner1998ecology, bronfenbrenner2007bioecological} argues that interactions between layers of systems influence the development of the individual. In these layers, the bioecological theory places an individual in the center of the \textit{microsystem}, surrounded with the individual’s immediate relationships, the \textit{mesosystem} captures interactions between levels or between different contexts in a given level, the \textit{exosystem} captures relationships with indirect environments, and the \textit{macrosystem} includes societal or belief level influences. These systems consist of elements that capture the \textit{process} (the interaction with objects or other people), \textit{person} (about the person, interest, appearance), \textit{context} (home, school, community, neighborhood), and \textit{time} (historical changes and the length of process). 
Although bioecological theory is commonly used in child development and family studies fields, it has also been applied in the process of designing technology for children as well, particularly in educational technology \cite{hartle2019technology}, e.g., designing telepresence robot interactions for supporting education of homebound children~\cite{ahumada2019going, ahumada2020theoretical}. Family Ecological Model (FEM) \cite{andrews1981ecological} is an adaptation of this bioecological theory, that similarly argues that family interactions, such as parenting, are shaped by the context that the families are surrounded, and families should be placed in the center of the ecology~\cite{davison2013reframing}.



Other ecological theories are applied in the design of new technology across HCI and HRI domains. 
`Product Service Ecology' model proposed by \citet{forlizzi2013product}, is an adaptation of `Social Ecological' model which allows designers to take a holistic approach to understanding the surrounding factors for the design and use of products. \citet{forlizzi2006service} applied an ecological model to study how autonomous service robots (i.e., Roomba vacuum cleaners)  fit in family homes, through ethnographic observations. Similarly, \citet{sung2010domestic} proposes the ``Domestic Robot Ecology'' model to support a holistic understanding of the complex in-home settings and acceptance of autonomous service robots for long-term use. An ``Ecology of Aging'' approach was applied to understand how robots can be designed to assist elderly at home \cite{forlizzi2004assistive}.  
However, these applications are typically focuses on domestic service robots, such as robotic cleaners, and there is limited research on ecological models for designing in-home robots that can afford \textit{social interactions, connections, and relationships} with children and families. 

\subsection{Designing for Family-Robot Interactions} 
Research in child-robot interaction has shown that children's interactions with robots at home can support their social and intellectual development. Robots motivate children to perform household chores like tidying up their rooms ~\cite{fink2014robot}, encourage children to read~\cite{michaelis2019supporting}, and support their emotion regulation~\cite{isbister2022design, slovak2018just}. However, user acceptance and long-term engagement with social robots still remains a challenge~\cite{de2016long, de2015living}. Social robots play different roles and provide companionship to their users~\cite{dautenhahn2005robot, cagiltay2020investigating}. Previous studies have highlighted a variety of robot roles for children, such as those played by robotic home assistants ~\cite{de2005assessing, dautenhahn2005robot}, socially assistive agents that provide autism~\cite{scassellati2018improving} and educational interventions ~\cite{belpaeme2018social}. Children and adolescents assign roles to social robots such as coaches, assistants, companions, and confidants~\cite{michaelis2018reading, cagiltay2020investigating, alves2022robots}, and elderly people open up to them about their frustrations~\cite{chang2015interaction}. 
Interactive social voice agents help children engage in social play and form positive bonds with robots through games and entertainment~\cite{belpaeme2012multimodal} and social robots can encourage family playtime and participation~\cite{kim2022can}. 

Co-design as a methodology can empower families to contribute to the design of new technologies through close collaboration with researchers. 
Including children and parents in design sessions reveals unique aspects of parent-child intimacy, i.e., increased involvement, affiliation, and sense of responsibility~\cite{dalsgaard2006mediated} and in-depth insight into family interactions in the design of novel technologies~\cite{yip2016evolution, cagiltay2020investigating}.
As family members may have conflicting goals and viewpoints, involving multi-generational family members in the design process can facilitate real-life use scenarios~\cite{christensen2019together} and the simple, flexible, and adaptable nature of technology probes can support family technology design and introduce playfulness to family dynamics~\cite{hutchinson2003technology}. 
A growing number of studies design in-home technologies with parents and children in mind~\cite{isola2012family, ahumada2019going, han2008comparative}. However, human-robot interaction research has a limited focus on families as a whole, i.e., family-robot interactions. To effectively design in-home social robots, it is essential to go beyond child-robot interactions and design for families as a whole, including children, siblings, parents, grandparents and extended family members or other household members. Overall, there is a gap in applying family-centered design approaches that captures the perspective of diverse family members in the design of in-home social robots.
\section{Family-Centered Child-Robot Interaction}

\begin{figure*}
\centering
  \includegraphics[width=\linewidth]{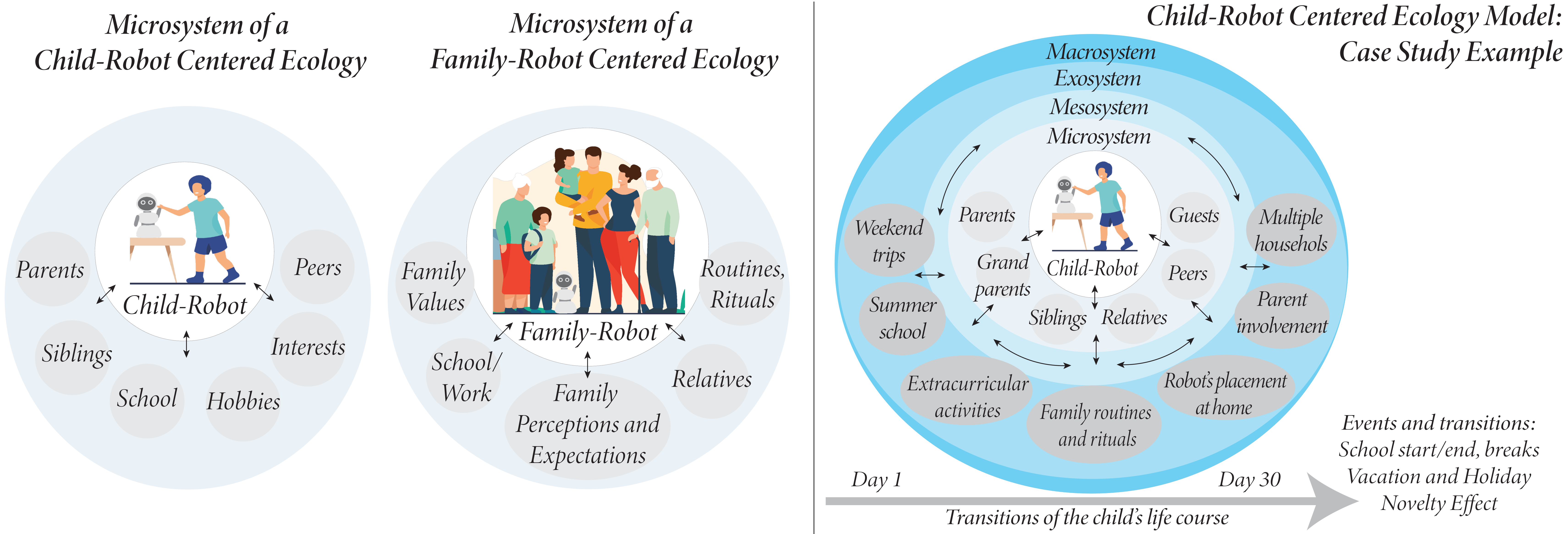}
  \Description{This figure has two panels and total of three illustrations of ecology models that surround child-robot or family-robot interactions. The left panel has two illustrations. The first illustration titled "microsystem of a child-robot centered ecology" represents a model where a child-robot pair is in the center. There are arrows pointing out to factors such as parents, siblings, school, hobbies, interests, peers. The second illustration titled "microsystem of a family-robot centered ecology" represents a model where a family-robot group is in the center. There are arrows pointing out to factors such as family values, school/work, family perceptions and expectations, relatives, routines and rituals. On the right panel titled "Child-Robot Centered Ecology Model: Case Study Example" there is a more high level illustration that captures different layers of the ecosystem, including macrosystem, exosystem, mesosystem, microsystem. The microsystem has the child-robot centered with arrows pointing to factors such as parents, grandparents, siblings, relatives, peers, guests. The mesosystem illustrates interactions between the systems through double-sided arrows, pointing within and between. The exosystem includes factors such as weekend trips, summer school, extracurricular activities, family routines and rituals, robot's placement at home, parent involvement, multiple households. Finally, on the bottom of the ecosystem, there is a timeline from day 1 to day 30, representing transitions of the child's life course, including events and transitions such as school start/end, breaks, vacation and holiday, novelty effect.}
  \caption{\textit{Illustration of Applications for the Family Ecological Model:} We illustrate the microsystem level of a child-robot and family-robot centered ecology (left). We illustrate the child-robot centered ecology example, described in Section 3.1 (right)}
  \label{FST}
\end{figure*}

Family ecological models place the family as a whole in the center of the ecology. From a family-centered child-robot interaction perspective, when we look at the ecology surrounding in-home social robots, we should place the family-robot subsystem as the center of the ecosystem. Past research in child-robot interaction however typically focuses on the child and robot dyad, and rarely considers the surrounding factors around the child and the robot. When designing social robots for children and families, focusing on the child-robot subsystem is a necessary baseline for a relationship to form in the first place, which can then afford to form family-level relationships in the home. For this purpose, we can either place the child-robot or family-robot at the center, depending on the affordances and purpose of the robot’s interaction design (See Fig~\ref{FST}). 
We will discuss two case studies that illustrate how family systems and family ecological factors contribute to children’s engagement with a robot, (1) from a child-robot centered perspective and (2) from a family-robot centered perspective. Within these case studies, we also demonstrate how the family theoretical models can inform the analysis and study design processes. 

\subsection{Long-Term Engagement with an In-Home Reading Companion Robot: Analysis}

We will demonstrate how using a family theoretical approach for analysis can inform our understanding how robots might be integrated in homes, from a child-robot centered perspective.

\textit{Context.} For this case example, we will take a recent study that designed a fully-autonomous reading companion robot and deployed it in 16 families' homes for one month~\cite{Cagiltay_engagement_2022}. This robot was designed to listen to the child as they read aloud an informal science book and make social commentary related to the book that aimed to support the child’s knowledge on the topic, interest in science, and form social connections through self-disclosure. It is notable that the robot only afforded dyadic interactions with the child, and was not designed to afford multi-party interactions. Findings from these unsupervised field deployments revealed different motivators for children's long-term engagement with the robot. Children's interest changed in the reading activity over time, and environmental factors such as parental influence (e.g., hands-off parents vs parents that set goals for their child or gave reminders), the immediacy of robot placement in the home (e.g., child’s bedroom or shared space and living room), changes to family routines (e.g., weekend trips, starting a new extracurricular activity, guest visits), and how well the robot activities conformed to those daily routines influenced children's engagement, interest, and adoption of the technology. Findings report that although many children in the one-month deployment adopted the robot, some families felt limited by a single type of interaction and had difficulties in integrating the robot into their routines which challenged their long-term engagement. Some children adapted and changed how they interacted with the robot based on their personal preferences, some were interrupted by external factors, and some discontinued the use of the robot. A secondary analysis of the daily interaction videos of families and explored how group interactions formed with and around the robot~\cite{offscript_HRI} and observed that many family members (e.g., parents, siblings, visiting relatives and guests) seemed eager to participate in group interactions but were unable to as the robot could only afford dyadic interactions. These shortcomings highlight missed opportunities to support interactions between family micro-systems (e.g., child-sibling-robot, parent-child-robot) which motivate the need for more family-centered approaches to facilitate family group interactions, described in this paper. 

\subsubsection{Applying Family Theories to inform analysis of child-robot interaction:} Given this context of the case study, we will take a deeper look into how a social robot that is a part of a family household impacts the behaviors between different family members, and vice-a-versa, from a FST and FEM perspective. Figure~\ref{FST} (right) illustrates the systems that surround the child-robot interactions in the aforementioned research context. From a family ecological approach, we place the child-robot dyad in the center due to the affordances of the robot interaction. The \textit{microsystem} around the child and robot captures the influential interactions and relationships that form with and around the robot. This microsystem includes family members, such as parents, grandparents, peers, siblings, relatives, guests. 
Interactions formed in this microsystem did not only include the child and the robot, but other dyadic interactions such as sibling-robot and guest-robot interactions, i.e., other family systems. Although the robot could not afford interactions with multiple group members, the video analysis from \cite{offscript_HRI} reported instances of multiple family members either passively observing or being directly involved in the child-robot interaction. 
%
The \textit{mesosystem} captures all possible interactions between systems in different contexts, for example how the robot's placement in the living room, a rule placed by a parent, might make it harder for the child to interact with the robot when the TV is on in the background, and cause conflict between the child and parent.
The \textit{exosystem} captures environmental factors that indirectly impact interactions between the child and robot, such as school or extra curricular activities, weekend trips, family routines and traditions, parent’s involvement in the study, and placement of the robot in the home. 
Furthermore, factors such as exposure to robot portrayals on media, recent developments on social robotics, or parents' relationship status are captured in the exosystem. For example, two participants in the study had multiple households, e.g., grandparent or co-parenting, where children split different amounts of their week across households throughout the study. This aspect of family structures indirectly impacted the child-robot subsystem as the child had to move the robot to different households to be able to keep interacting with it. Over time these children were observed to start leaving the robot in one of their households, and did not prefer to move the robot to the other. Children's motivation behind this observation were not explored within the scope of the study, however this example demonstrates that not only family ecology but also family-structures impact interactions between the child and robot. 
 \textit{Macrosystem} level factors can include cultural attitudes and ideologies towards robots, which were not captured within the scope of the study by~\citet{Cagiltay_engagement_2022}.
\textit{Temporal changes} over time in the child’s life course, such as life events, transitions, and the novelty effect of the robot diminishing over time are factors that impact the child’s engagement with the robot in the long-term. 
Overall, this case exemplifies that family is an integral part of the child-robot ecology and cannot be separated from the microsystem, thus is needed to be considered in robot design decisions. Family theories FEM and FST can be useful to help identify and draw attention to potential influences on child-robot interactions that impact the adoption, use, or disengagement of robots within a family. 
%

\subsection{Family-Robot Interactions at Home: How can Family Theories Inform Study Design?}
%
In this second case example, we will discuss how family theories, FST and FEM, can be useful in a study design process to inform research in exploring how robots might be integrated in homes, from a family-robot centered perspective.
We specifically focus on the context of fully autonomous in-home family-robot interactions \textit{(i.e., multiple family members interacting with a social robot together with a shared purpose)}, which is not commonly explored in child-robot interaction research.
We present a hypothetical study design to capture the complexity of designing social robot interactions for children and their families, and discuss how a family theoretical approach can help this design process.

\textit{Context.} Consider the following context where a social robot is designed to facilitate family connections and social interactions in shared recreational activities, for example playing the game of Charades or a social deduction board game. The study aims to explore changes in families’ long-term interactions with this robot. The robot’s role in this case is to be an active participant in the game and facilitate social connections between family members through expressing comments about the game, describing rules, making jokes, facilitating conversations, managing conflict etc. Any family member can initiate the interaction with the robot at any preferred time. The robot will be deployed in 20 family homes with at least one child aged 8-16, for 6-months. Collected data will include interaction logs, monthly interviews, and feedback surveys. 
Below are a few specific ways in which family theories, FST and FEM, can inform the study design: 

\subsubsection{A family-centered study design to inform family-robot interaction design} Family-centered study design could inform research questions, metrics for data collection, and sample selection. 
Firstly, FST and FEM can be used to structure \textit{research questions} that are relevant to studying family-robot interactions. For example, researchers might ask questions about how the robot is integrated into the family’s routines and activities, how the relationships between family members and the robot evolve over time, and how the robot's presence changes the dynamics of family interactions, such as who initiates conversations and how authority and priority conflicts are managed within the family. 
Secondly, FST and FEM can be used to identify the most relevant aspects of family-robot interactions to be captured in the \textit{data collection process}, which can then inform necessary \textit{metrics} to better understand family-robot interactions. For example, FST can inform the selection of measures to assess family communication and relationship quality, while FEM can inform the selection of measures for which family members and surrounding factors to assess.
The robot’s interaction log can be structured to report the frequency of families interacting with the robot, which family members in the household are involved in the interactions, and what types of family interactions form with and around the robot. Qualitative interview protocols can be designed to include questions capturing robot perceptions and relationships of different family subsystems.
Thirdly, FST and FEM can also inform the study's \textit{sample selection}. For example, the study may call for the need to recruit families with different structures (e.g.,~\cite{pasley2016family} binuclear, blended, cohabiting, multigenerational families etc.) to explore if and how the robot's interactions are differently perceived in different types of family structures.

\subsubsection{Family Theories to interpret family-centered study findings} FST and FEM can inform the interpretation of the study's findings. For example, for a robot that is designed to facilitate conversations and connection between family members, \textit{FST} can identify the robot's relationships between different family subsystems like parent-child, sibling, spouses and help understand which family subsystems are more likely to initiate conversations with or respond to the robot. If the study findings show that family members have different perceptions or expectations towards the robot, FST can help understand how these perceptions could inform design decisions for robot interactions by taking these subsystems into account. 
Study findings can be interpreted using \textit{FEM} to understand ecological factors that may affect robot adoption and use, such as: \textit{microsystem} factors that focus on the family-robot system and their immediate environment; \textit{mesosystem} level factors that capture interactions between family systems, their values, beliefs, and robot perceptions and expectations; \textit{exosystem} level factors that capture family exposure to media portrayals of robots; and \textit{macrosystem} factors that capture the family's socioeconomic status, access to resources and support, political legislation, cultural norms and expectations about robot use.
Informed by study findings, FEM and FST can support a comprehensive understanding of how different factors impact families' robot adoption and use at homes.

\section{Discussion}
%
Our question motivating this work was: \textit{How can family theories, i.e., Family Systems Theory and Family Ecological Model, inform how robots might be integrated in domestic environments to support child-robot and family-robot interaction?} Through two case studies, we demonstrated how FST and FEM could inform design decisions in child-robot and family-robot interaction research. FST is a useful theory to understand how different family members can affect interactions with the social robot at home. FEM as a framework helps understand how different ecological factors can impact family interactions with social robots at home.
Families are complex and diverse, and thus it is important for future research in social robotics and child-robot interactions to pursue a family-centered approach. In this section, we discuss the strengths and limitations of these theories, followed up by recommendations for a family-centered design approach in child-robot interaction research.

\subsection{Strengths and Limitations of Family Theories in Child-Robot Interactions}

Family theories FST and FEM allow us to center ourselves around a child-robot and family-robot subsystem. From this center, we can analyze the relationships and behaviors surrounding the family systems that either affect or are affected by the child-robot subsystem. We can also understand relationships that extend the child-robot system, such as the sibling-robot, or parent-child-robot, parent-sibling-child-robot systems. One limitation of the FST approach is that it does not consider the temporal aspects of interactions. For example, events and transitions of a child’s life course are not captured as a key part of FST. However, even for the relatively short period of four weeks in the study discussed earlier \citet{Cagiltay_engagement_2022, offscript_HRI}, findings captured events such as summer school, overnight visits from guests and relatives, or weekend travels as factors that influenced the interactions within the child-robot subsystem. In social robotics research, the novelty effect suggests that people's initial responses to a robot starts the strongest, resulting in an overly positive or negative view of robots that fades over time. Longer-term research studies (six months or more) with families may be affected by temporal factors such as the novelty effect, events, and transitions in the child's life.
Because FST does not capture the temporal aspects of interaction, we addressed this limitation in the first case study by incorporating the \textit{time} element from Bronfenbrenner’s bioecological theory ~\cite{bronfenbrenner1979ecology, bronfenbrenner1998ecology, bronfenbrenner2007bioecological} into our family ecological model.

A strength of these theoretical models is their flexibility and adaptability to be applied in any context under child-robot and family-robot interaction research. Such contexts can include research in lab or home settings, as well as different cultural contexts, age groups, and variety of family structures. 
This adaptability is an advantage because, just as families do, robots come in various shapes and forms, with varying technical and social affordances. 
However, despite the significant recent technological advancements in the robotics field, it is still technically challenging for robots to process complex information in dynamic real-world environments, like homes, which might affect the robot's ability to form basic social interactions with people. 
These technical challenges are especially prevalent in multi-party interactions. For example, today's commercial social robots come with the ability to detect the number of people in the room, make eye-contact with addressees, and understand basic body gestures and expressions of emotion. However, more socially complex features, such as understanding relationship levels between people, understanding the dynamic changes in these relationships, effectively recognizing and responding to human emotions, accurately detecting and processing speech of children or elderly, and holding meaningful and natural conversations with them, remains open challenges. Therefore, further technological advancement is needed to enable social these robots to effectively interact with people in complex family environments. 
The flexibility of FST and FEM frameworks offer a great advantage in supporting research in family-centered human-robot interaction. 

\subsection{Recommendations for Family-Centered Design in Child-Robot Interaction Research} 

The case studies establish that when designing child-robot interactions for the home, it is likely that many family members will be involved and invested in the interaction, and ecological factors will impact interactions not only from the microsystem level but also in the meso, exo, and macro level. Capturing these multifaceted factors can help guide researchers to better understand the integration of robots in the home.
Below, we discuss important recommendations for using family theories when designing for in-home child-robot interactions with a family-centered approach.

\subsubsection{(R1) Shift focus from the child-robot dyad to the family as a whole}
Taking a FST perspective argues that anyone surrounding the child and robot could influence the interaction or be a part of the interaction. When investigating the ecological factors, place the child-robot or family-robot at the center (as seen in Figure \ref{FST}), depending on the affordances and purpose of the robot’s interaction design. 

\textit{Child-Robot in the center of the ecosystem} assumes the child is the primary actor with the robot, while multiple household family members could still interact with or be present around the robot (closely connected), influencing the child-robot interaction. This approach designs for the child first, but still considers the family part of the ecology.

\textit{Family-Robot in the center of the ecosystem} affords both dyadic and multi-party interactions with dynamic changes in group formations. Placing the family in the middle assumes family members can interact with the robot as a group, or individually. This could potentially include interactions with the robot where the child of the household is not involved. Group members can influence each others’ interaction with the robot, and the robot can help facilitate connections between members. This approach designs for the child and family as a whole.

\subsubsection{(R2) Understand families' expectations, concerns, and acceptance towards robots: From a child-centered and family-centered perspective}
FST can help identify parental expectations, concerns, and acceptance of social robots in the home~\cite{lin2020parental}. In a \textit{child-centered ecosystem}, although family members will have preferences about the robot, the primary focus should be the child’s expectations, concerns, and long-term acceptance. However, in a \textit{family-centered ecosystem}, all members of the family household will have direct stakes in the interaction, thus their expectations, concerns and acceptance of the robot will directly influence the child’s and the family’s interactions with the robot. It is likely that conflicting opinions and changes in expectations or concerns toward the robot will impact interactions with the robot more in a family-centered ecosystem, compared to a child-centered ecosystem.

\subsubsection{(R3) Employ family ecological model as a framework for study design}
We demonstrated a case for how FEM could be a useful framework for making sense of child-robot interaction data and help shape study design processes. Proactively preparing hypothetical or theoretically grounded ecological models for the family-robot interaction could be a valuable process in determining the materials that will be necessary to measure environmental factors outside of the microsystem (i.e., meso, exo, macro system factors). Prepare to capture the complex and diverse nature of families and ecological factors that impact their engagement with the robot. An extension of this design recommendation is described in R4.

\subsubsection{(R4) Design for and adapt study designs to different family structures}
The diversity in family structures should be represented in research. Similar to our argument in R1, such that designing solely for the child is insufficient, designing only for one family structure is insufficient when tailoring social robot interactions for families. Diverse family structures could range from nuclear families to multi-household, multi-generational families and from single-parent to binuclear families~\cite{pasley2016family}. Plan for adjustments by preparing versions of study protocols to accommodate diversity in families. Some examples could include transportation protocols, i.e., if participants need to frequently move between different family homes, or if participants are not able to move the robot to another household with them. Design study protocols considering families with accessibility needs (e.g., elderly or disabled family members). Instead of screening out particular family structures or being unprepared for these structures, this practice will allow to include diverse families and gain a richer understanding of families and their interactions with robots. Though this design implication might be challenging to execute and is highly context dependent, its application can provide a unique and inclusive understandings to the design of technology.

\subsubsection{(R5) Consider temporal changes, events, and transitions when designing for family-centered child-robot interactions}
Events, transitions, and changes are inevitable in a developing child’s life. Changes related to school, family, friendships, interests, and growth happen over time, and these changes may look different depending on the child's age, time of the year, and type of influence children get from their environment (people, media, etc). From an FST perspective, any change for one member will also affect other members of the family to an extent. Thus, changes don't only happen in isolation within the child's life but are inevitable for families as a whole. Temporal changes (unexpected or expected, short or long term) can occur in family members' employment or health; legislation relevant to families, such as taxation, schooling, etc.; or family structures (addition/loss of a family member). From a robot designer's perspective, such changes, events, and transitions will affect families’ engagement with the robot, the frequency of interactions with the robot, or the effectiveness of the robot’s social intervention. Therefore, it is beneficial to capture these temporal aspects to the best of our ability via targeted qualitative research methods or quantitative survey metrics integrated into study design practices. In the analysis, capturing the emerging temporal factors within the ecosystem also allows us to make sense of the data by considering all these changes. If designed effectively, future design opportunities for social robots include supporting families in moments of crisis or change, in maintaining routines and rituals, and in developing resilience against changes.

\section{Conclusion}
A theoretically motivated family-centered design approach to child-robot interactions is critical to better understand long-term engagement and acceptance of robots in the home. To demonstrate the complexity and importance of such approaches, we presented two case studies that exemplified how family systems theory and family ecological model are relevant and can be applied to child-robot interaction design. We argue that family theories can be insightful in both the analysis and study design processes. Finally, we discussed the benefits and limitations of employing family theoretical perspectives, and offered design recommendations for child-robot interaction researchers. Overall, family theories can be useful frameworks for understanding and analyzing the complex dynamics of human-robot interaction within a family context, and can inform how robots might be integrated in homes to support family and child robot interactions. 

\begin{acks}
This work was partially supported by National Science Foundation award \#2202802.
\end{acks}

\section{Selection and Participation of Children}
No children participated in this work. This work discusses theoretical framework for a family systems and family ecological approaches. The case study included in the analysis reflects on data from prior published articles in child-robot interaction field~\cite{Cagiltay_engagement_2022}, which included details on selection and participation of children.

\bibliographystyle{ACM-Reference-Format}
\bibliography{manuscript}

\end{document}